\documentclass[showpacs,prb,floatfix,superscriptaddress,twocolumn,amsmath]{revtex4-1}
\usepackage[T1]{fontenc}
\usepackage[latin1]{inputenc}
\usepackage[english]{babel}
\usepackage{epsfig}
\usepackage{soul,xcolor}
\usepackage{color}
\usepackage{textcomp}
\usepackage{amssymb}  
\usepackage{amsmath}  
\usepackage{latexsym} 
\usepackage{amsthm}   
\usepackage{soul}   
\newcommand{\be}{\begin{equation}}
\newcommand{\ee}{\end{equation}}
\newcommand{\bea}{\begin{eqnarray}}
\newcommand{\eea}{\end{eqnarray}}
\newcommand{\ba}{\begin{array}}
\newcommand{\ea}{\end{array}}

\definecolor{myGreen}{rgb}{0.1,0.6,0.1}

\newcommand{\qvec}{{\bf q}}   
\newcommand{\dd}{\text{d}}
\newcommand{\cmi}{{$\text{cm}^{-1}$}}
\newcommand{\texp}{\text{exp}}

\begin{document}
\title{ Glue function of optimally and overdoped cuprates from \\ inversion of the Raman spectra }  
\author{L. Fanfarillo}
\affiliation{Instituto de Ciencia de Materiales de Madrid, 
ICMM--CSIC, Cantoblanco, E--28049 Madrid, Spain}
\author{M. Mori}
\affiliation{Dipartimento di Fisica, Universit\`a di Roma Sapienza, 
Piazzale Aldo Moro 5, I--00185 Roma, Italy}
\affiliation{Departamento de Bioqu\'imica y Biolog\'ia Molecular I, Facultad de Ciencias Qu\'imicas, 
Universidad Complutense de Madrid, Ciudad Universitaria, E--28045 Madrid, Spain}
\author{M. Campetella}
\affiliation{Dipartimento di Chimica, Universit\`a di Roma Sapienza, 
Piazzale Aldo Moro 5, I--00185 Roma, Italy}
\author{M. Grilli}
\affiliation{Dipartimento di Fisica, Universit\`a di Roma Sapienza, 
Piazzale Aldo Moro 5, I--00185 Roma, Italy}
\affiliation{Istituto dei Sistemi Complessi CNR and CNISM Unit\`a di Roma Sapienza}
\author{S. Caprara}
\affiliation{Dipartimento di Fisica, Universit\`a di Roma Sapienza, 
Piazzale Aldo Moro 5, I--00185 Roma, Italy}
\affiliation{Istituto dei Sistemi Complessi CNR and CNISM Unit\`a di Roma Sapienza}

\begin{abstract}
We address the issue of identifying the mediators of effective interactions in 
cuprates superconductors. Specifically, we use 
inversion theory to analyze Raman spectra of optimally and over-doped 
La$_{2-x}$Sr$_x$CuO$_4$ samples. This allows us to extract the so-called glue 
function without making any {\it a priori} assumption based on any specific 
model. We use instead two different techniques, namely the singular value 
decomposition and a multi-rectangle decomposition. With both 
techniques we find consistent results showing that: i) two distinct excitations 
are responsible for the glue function, which have completely different doping 
dependence. One excitation becomes weak above optimal doping, where 
on the contrary the other keeps (or even slightly increases) its strength; ii) 
there is a marked temperature dependence on the weight and spectral distribution 
of these excitations, which therefore must have a somewhat critical character. 
It is quite natural to identify and characterize these 
two distinct excitations as damped antiferromagnetic spin  waves and damped 
charge density waves, respectively. This sets the stage for a scenario in which 
superconductivity is concomitant and competing with a charge ordering 
instability.
\end{abstract} 

\date{\today} 
\maketitle

\section{Introduction}
The issue of the mediators of high-temperature superconductivity in cuprates, 
the so-called ``glue issue'', is still unsolved. On the one hand, there is the 
early proposal of an essentially instantaneous magnetically mediated pairing 
forming incoherent singlets eventually condensing in the superconducting 
state.\cite{glue} On the other hand, it has been proposed that slow, nearly 
critical collective excitations mediate a retarded pairing. In this latter 
framework overdamped spin waves are a natural candidate due to the proximity of 
the superconducting phase to the Mott insulating antiferromagnetic 
phase.\cite{chubukov,dahm} How strong is the instantaneous or the retarded 
character of the magnetic superconducting glue is likely  just a matter of 
quantitative balance.\cite{scalapino,hanke} On the contrary, there are other 
proposals suggesting that the superconducting mediator could also have a 
non-magnetic character. In this regard, primary candidates are charge-ordering 
fluctuations, which were proposed as a source of retarded glue already long 
ago.\cite{GRCDK,perali,CDGZP} This last proposal has steadily acquired 
importance with the progressive increasing experimental evidence of charge 
ordering as a common feature of cuprates via neutron scattering,\cite{tranquada}
NMR and NQR,\cite{jullien} EXAFS,\cite{bianconi} ARPES,\cite{Seibold,mazza,hashimoto}
STM,\cite{howald,yazdani,hanaguri,kohsaka,jlee} and Raman 
scattering,\cite{tassini,suppa,Caprara2011,colonna} until it has become a compelling evidence 
due to the recent direct observation by resonant X-ray 
scattering.\cite{abbamonte,ghiringhelli,chang,dasilva,comin,letacon}

The evidence of charge ordering and the simultaneous natural presence of spin 
waves emanating from the nearby AFM state raises the issue of coexistence, 
interplay, and possible causal relationship between charge and spin degrees of 
freedom. An early point of view is that charge ordering arises from 
Coulomb-frustrated  charge segregation due to short-range (essentially 
instantaneous) magnetic interactions\cite{emerykivelson,CGK} or/and due to 
non-magnetic (e.g., phononic\cite{CDG,BTGD}, charge-transfer\cite{CT1991},...) attractive forces. 
In both cases, the concomitant presence of charge-ordering fluctuations ``enslaves''
the spin degrees of freedom, allowing them to survive up to large doping. This tight 
entangling of spin and charge degrees of freedom is also an intrinsic feature of 
the `stripe'' concept.\cite{noistripes} A complementary, more recent, view is that collective 
retarded spin fluctuations may give rise to attraction in the particle-hole 
channel inducing a charge-density wave instability.\cite{metliski,wangchubukov}
At the moment, it is unclear and debated whether these two points of view are 
really distinct or just some kind of ``egg-and-chicken'' issue arising from an 
underlying continuous interplay between charge and spin degrees of freedom. In 
any case it seems by now rather natural to assume that charge-density waves and 
spin-density waves coexist and it is quite interesting to investigate the 
relative weight of these different excitations in mediating the effective 
electron-electron interaction in cuprates in the  different doping and 
temperature regimes. This is precisely the focus of the present work.

The presence of two distinct sources of glue in cuprates was found both in 
optical\cite{vanheumen,dalconte}, ARPES\cite{schachinger, he_prl13} and 
Raman\cite{Muschler2010,Caprara2011} experiments. These experiments 
identify the so-called ``glue function'' $\alpha^2F(\omega)$, introduced by 
Eliashberg to characterize the spectral distribution and the strength of the 
electron-phonon coupling in the superconducting pairing.\cite{eliashberg} From 
optical conductivity it is found that the glue function has a double structure 
with a rather narrow and temperature dependent peak at low frequencies (below 
$10^3$\,cm$^{-1}$) and a broad structure extending up $3000-4000$\,cm$^{-1}$. 
From the characteristic frequency ranges it is quite natural to attribute the 
first peak to phononic excitations, while the mediators responsible for the 
broad peak should be diffusive magnetic excitations.\cite{normanchubukov} 
However, neutron scattering experiments show that the magnetic excitations have 
a rather rich spectral structure,\cite{wakimoto1,wakimoto2} which, besides the 
broad high-frequency peak, also display a low-frequency peak. Moreover, 
time-resolved optical spectroscopy finds that also the low-frequency peak could 
be related to electronic excitations.\cite{dalconte} Thus the 
attribution and identification of the various excitations responsible for the 
effective electron-electron interaction is still an open relevant issue. In this 
framework, the analysis of Raman spectra is interesting because this technique 
is able to explore selectively different regions of the electronic Brillouin 
zone allowing to extract more detailed information on the electronic excitations 
than the optical conductivity. This selectivity of Raman spectra was already 
exploited in Ref. \onlinecite{Caprara2011} to identify and characterize the 
double nature of the electronic mediators, but the analysis was carried out 
fitting the experiments by analytic expressions arising from a specific 
assumption on the form of the charge and spin collective excitations. In 
particular the standard diffusive form of nearly critical overdamped collective 
modes near gaussian quantum critical points was assumed. This not only allowed a 
good fitting of the spectra, but allowed to attribute the markedly different 
shape of the Raman spectra in the different channel (the $B_{1g}$ and $B_{2g}$ 
scattering channels  obtained by differently polarizing the incoming and 
outgoing photons of the Raman scattering) to different, spin and charge, 
excitations, which acted differently on the spectra according to their different 
characteristic wavevectors. Although successful, this approach was starting from 
specific assumptions on the form of the mediators, which had the single spectral 
structure of overdamped modes. It is therefore quite important to test and 
deepen the conclusions of Ref.\,\onlinecite{Caprara2011} with a more flexible and 
general approach. In this work we apply two different numerical
inversion techniques to extract the glue function from the Raman spectra and we 
draw consistent conclusions: i) two distinct excitations are responsible 
for the glue function, which have completely different doping dependence; ii) 
there is a marked temperature dependence of the weight and spectral distribution 
of these excitations, which therefore must have a somewhat critical character. 
Indeed, if they were just the result of low-energy phonon-mediated interaction 
and high-energy local excitations due to strong-correlation effects, any 
temperature dependence  would hardly occur. 

Our analysis is purposely carried out on optimally and overdoped samples to 
highlight the specific features of the bosonic interaction mediators only. This analysis could
of course be extended to the underdoped regime, where, however, the presence
of a pseudogap in the fermionic spectrum introduces additional temperature, doping, and
momentum dependencies, which would be superimposed to those of the bosonic mediators.
This would make the interpretation of the data much less conclusive and informative.

The paper is organized as follows: In Section II we describe the two numerical 
methods adopted to extract the glue function from the Raman spectra. Sections 
III and IV contain the results of our analysis of the glue functions and of their 
low-frequency behavior. Finally, 
Section V reports and discusses our concluding remarks. Appendices A and B 
contain technical details on the implementation of the inversion procedure,
while Appendix C contains details on the extraction of the low-frequency 
properties of the glue function.

\section{Glue function and Raman spectra}
In this work we analyze Raman spectra from Ref.\,\onlinecite{Caprara2011} in the 
$B_{1g}$ and $B_{2g}$ channels in La$_{2-x}$Sr$_x$CuO$_4$ (LSCO) samples for 
different temperatures and doping levels. Electronic Raman scattering is a bulk 
(nearly surface-insensitive) probe and it measures a response function $\chi(z)$ 
analogous to that of the optical conductivity. The Raman response function can 
be expressed\cite{goetze1972_memfun} as a function of the so-called memory 
function $M(z)$ as $\chi(z)=\chi_0 M(z)/[z+M(z)]$ for complex arguments $z$, 
where $\chi_0$ is the (real) ``bare''  response function computed in absence of 
any scattering process. The imaginary part of the response function, $\chi''$, 
can be written as:
\begin{equation} \label{eq:im_chi_and_memfun}
\frac{\chi''(\omega)}{\chi_0} = \frac{  \omega M''(\omega)}{[M'(\omega)+\omega]^2 + [M''(\omega)]^2}.
\end{equation}
Here, the real and imaginary parts of the memory functions, $M(\omega) = 
M'(\omega)+iM''(\omega)$, are function of the real frequency $\omega$, and they 
are related by a Kramers-Kronig (KK) transformation. In turn, $M''(\omega)$ can 
be expressed\cite{eliashberg} in terms of the glue function $\alpha^2 F(z)$ by 
means of the integral expression
\begin{equation} \label{eq:integralEquation}
M''(\omega) = \int_0^\infty \dd z\; K(\omega,z) \alpha^2 F(z),
\end{equation}
with the kernel
\begin{align} \label{eq:kernelDefinition}
 K(\omega,z) = \frac \pi \omega &\left[2 \omega \coth \left( \frac{z}{2T} \right) \right. 
               - (z+\omega)\coth \left(\frac{z+\omega}{2T} \right)+ \\ &
             \left.  (z-\omega) \coth \left(\frac{z-\omega}{2T} \right) \right]\,. \nonumber
\end{align}
Therefore, $\chi''$ is a non linear functional of $\alpha^2F$. The extraction of 
$\alpha^2F$ from experimental data is possible by approximating it as a linear 
combination of suitable basis functions as $\alpha^2F(\omega) \approx 
\sum_{\alpha=1}^N c_\alpha \phi_\alpha(\omega)$. Once a basis is chosen, one may 
optimize the coefficients $c_\alpha$ so as to fit the experimental data. 
In this way, Eq.~\eqref{eq:integralEquation} can  be rewritten as
$M''(\omega) = \sum_\alpha c_\alpha A''_\alpha (\omega)$,
with $A''_\alpha(\omega) = \int_0^\infty \dd z \; K(\omega,z) \phi_\alpha(z)$,
so that Eq.~\eqref{eq:im_chi_and_memfun} becomes 
\begin{equation} 
\label{eq:im_chi_and_memfun_expanded}
\frac{\chi''(\omega,\{c_\alpha \})}{\chi_0} = \frac{ \sum_{\alpha}  c_\alpha \omega 
A_\alpha ''(\omega)}{[\sum_{\alpha} c_\alpha A_\alpha'(\omega)+\omega]^2 + 
[\sum_{\alpha} c_\alpha A_\alpha''(\omega)]^2}\,.
\end{equation}
where $A'_\alpha(\omega)$ is the KK transform of $A''_\alpha(\omega)$.
The coefficients $c_\alpha$ are then used as fit parameters. We used and compare 
the results obtained with two very different choices for the  basis functions, 
the first one based on a multi-rectangle decomposition (MRD) of the glue function, and 
the second based on the singular vector expansion of the kernel $K$. 
\begin{itemize}
\item {\bf Multi-Rectangle Decomposition (MRD)}. 
The glue function is approximated by a piecewise constant function, 
which corresponds to choosing a partition of the frequency axis 
$\omega_1<...<\omega_\alpha<...<\omega_{N+1}$ and $N$ non-overlapping box 
functions $\phi_\alpha(\omega)=1$ for $\omega_\alpha<\omega<\omega_{\alpha+1}$ 
and $\phi_\alpha(\omega)=0$ elsewhere,  as the basis functions. The fitting 
parameters $c_\alpha$ are the heights of the box functions. In the MRD approach the 
functions $A''_\alpha$ can then be computed analytically 
(see Appendix~\ref{sect:binning_appendix} for further details).
\item {\bf Singular Vector Decomposition (SVD)}. 
The starting point of this approach is the expansion of the integral kernel 
Eq.~\eqref{eq:integralEquation} as a sum of diadic operators:
\[
K(\omega,z) = \sum_{\alpha=1}^\infty \sigma_\alpha \psi_\alpha(\omega) \phi_\alpha(z)
\]
where the singular values $\sigma_\alpha$ are nonnegative and in decreasing 
order, and $\{\psi_\alpha \}$ and $\{\phi_\alpha \}$ are sets of orthogonal 
functions, called the left and right singular vectors, respectively. In this 
case we approximate the glue function using as basis the first $N$ 
right-singular vectors of the singular vector decomposition of the kernel 
\cite{Christian2010_SVDbook,Martin2012_svd}. This allows us to define 
$A''_\alpha(\omega)=\sigma_\alpha \psi_\alpha(\omega)$. Notice that the integral 
kernel of our physical problem has to be properly regularized before the 
expansion thus leading to slightly different expressions; we refer to the 
Appendix~\ref{sect:SVD_appendix} for a detailed derivation.
\end{itemize}

In both cases coefficients $c_\alpha$ are computed by minimizing the square distance $\Delta^2$ between the 
theoretical response function and the experimental one. In order to do this, the functions $A'$ and $A''$ are 
computed at the experimental points $\omega_i$, $i=1,\dots,N_{\text{exp}}$. We then look for
\begin{equation} \label{eq:Delta2}
\min_{\{c_\alpha\}}\Delta^2(\{c_\alpha\}) \equiv \min_{\{c_\alpha\}} \sum_{i=1}^{N_\texp} \left[\chi''_{\texp,i} - 
\chi''_{\text{th},i}(\{c_\alpha\}) \right]^2,
\end{equation}
where $\chi''_{\text{th},i}(\{c_\alpha\})\equiv\chi''(\omega_i,\{c_\alpha \})$ is computed according to 
Eq.\,\eqref{eq:im_chi_and_memfun_expanded}, and $\chi''_{\texp,i}\equiv\chi''_{\texp}(\omega_i)$ are the 
measured values. The minimization is constrained by the requirement $\alpha^2 F(z) \ge 0$, 
which can be written in the MRD case as $c_\alpha \ge 0$, and in SVD case as 
$\sum_\alpha c_\alpha \phi_\alpha(z)\ge 0$, for all possible values of $z$.
Further details about the choice of the functions $\phi_\alpha$ and $\chi_0$, and in the discretization 
and in the fitting procedures can be found in Appendices~\ref{sect:binning_appendix} (MRD) and 
\ref{sect:SVD_appendix} (SVD).

\section{$B_{1g}$ and $B_{2g}$ Glue Functions: Doping and temperature dependence}
\label{sect:fit_results}
\begin{figure*}
\leavevmode
\begin{center}
\hspace{10.0truecm}
\includegraphics[width=0.95\textwidth]{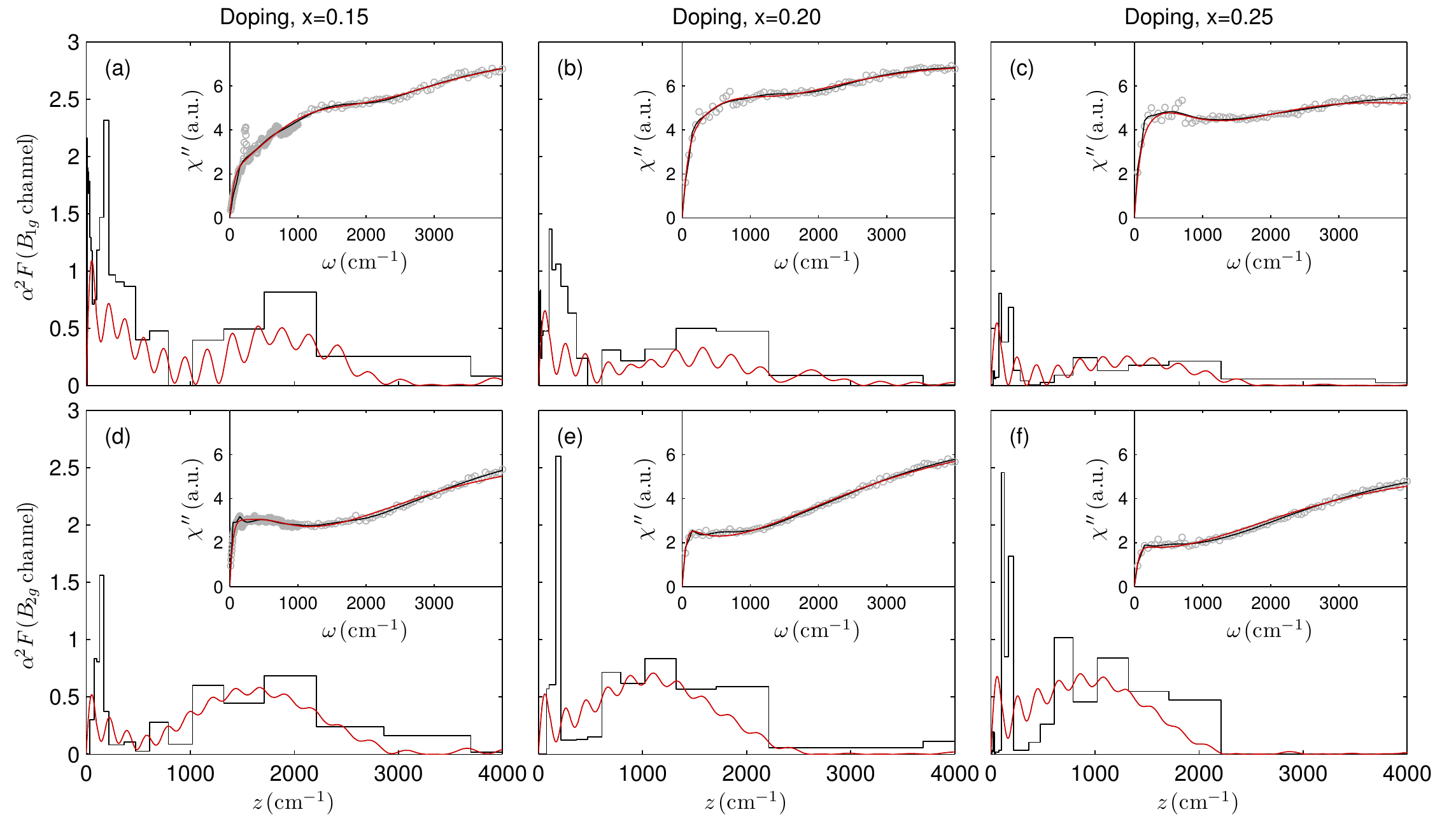}
\caption{(Color online)Fitted glue functions $\alpha^2F(z)$ using the SVD (red) and MRD (black) methods, for three different 
doping levels [(a) and (d), $x=0.15$; (b) and (d), $x=0.20$; (c) and (f), $x=0.25$], for both $B_{1g}$ [(a)--(c)] and $B_{2g}$ 
[(d)--(f)] channels at low temperature ($\sim 50$\,K). The insets show the experimental data (grey circles) and 
the response functions computed from the fitted glue functions.}
\label{fig:figure1}
\end{center}
\end{figure*}

We analyze Raman spectra of La$_{2-x}$Sr$_x$CuO$_4$ samples focusing on
three different doping levels ($x=0.15,\, 0.20,\, 0.25$) and three different temperatures 
(approximately 50, 100 and 200\,K; the true temperatures for different doping levels are slightly different), to highlight 
the doping and temperature evolution. 
In Fig.\,\ref{fig:figure1} we show the glue functions obtained with both MRD and SVD techniques by analyzing data
of three different doped samples ($x=0.15,\, 0.20,\, 0.25$) at $T\sim50$\,K, for both the $B_{1g}$ and $B_{2g}$ channels.

The glue function obtained by SVD has an oscillatory character, due to the fact that it is obtained by summing up 
oscillating functions with increasing frequency, similarly to a Fourier expansion. (See Fig.~\ref{fig:suppfig_SVD_functions}
for representative examples of the basis functions.) On the other hand, the MRD 
approach may yield very irregular glue function. Some sort of indeterminacy is an inevitable consequence of the 
structure of the problem. Integral kernels act as low-pass filters, and therefore only the low-frequency components 
of the glue function, related to the largest singular values, can be reliably determined from the (noisy) data. 
(See Fig.~\ref{fig:rank} for the magnitude of the singular values of the kernel.) 
Instead of relying on arbitrary smoothing techniques, we proceeded in a pragmatic twofold way. On the one hand
we considered integrated quantities, 
which are relatively insensitive to both discontinuities and oscillations in the glue functions, and therefore allow 
to extract robust relevant informations.
On the other hand, one can notice that despite the different ``looks'' of the glue functions obtained with the two approaches,
some features are common and therefore more reliable.
Indeed, both approaches yield in both
channels two structures:  a low-frequency peak ($\omega \lesssim 
500$\,\cmi) and a broader structure in the range 500--3000\,\cmi. The $B_{1g}$ 
channel appears to be suppressed as the doping is increased, while the secondary 
structure in the $B_{2g}$ glue function shifts to lower frequencies as the 
doping is increased. These behaviors are also present at $T=100$ and 200\,K. 
This latter case is reported in Fig.~\ref{fig:figure1b}. Due to the 
behavior of the kernel at high temperature (see discussion in 
Appendix~\ref{sect:SVD_appendix_fitting}),  the structures of the glue 
functions are rougher, but still the main features found at $T=50$\,K persist, 
namely a generic two-peak structure and a different doping dependence of the two 
channels.

\begin{figure*}
\leavevmode
\begin{center}
\hspace{10.0truecm}
\includegraphics[width=0.95\textwidth]{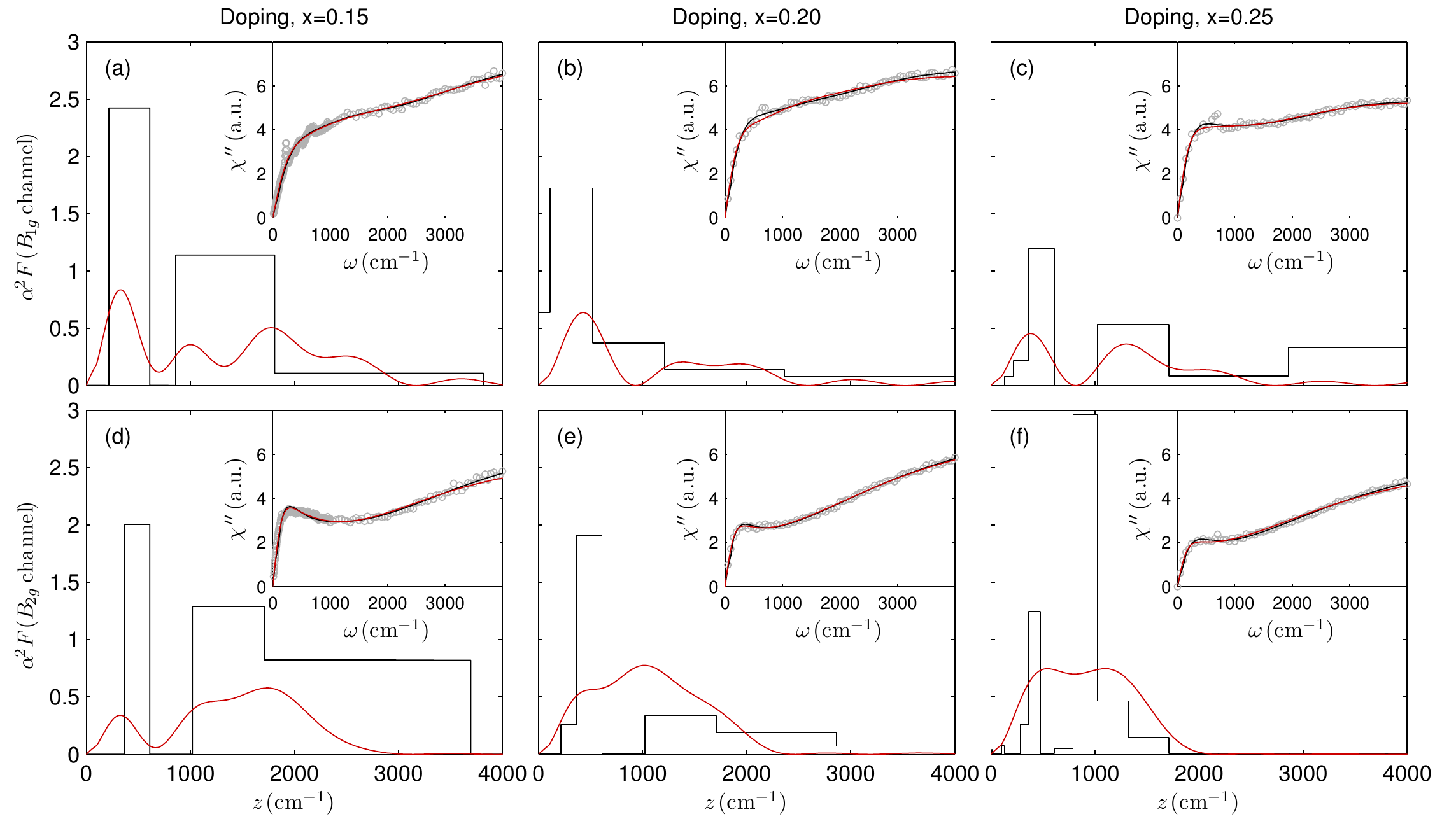}
\caption{(Color online) Fitted glue functions $\alpha^2F(z)$ using the SVD (red) and MRD (black) methods, for three different 
doping levels [(a) and (d), $x=0.15$; (b) and (d), $x=0.20$; (c) and (f), $x=0.25$)], for both $B_{1g}$ [(a)--(c)] and 
$B_{2g}$ [(d)--(f)] channels at high temperature ($\sim 200$\,K). The insets show the experimental data (grey circles) 
and the response functions computed from the fitted glue functions.}
\label{fig:figure1b}
\end{center}
\end{figure*}

\begin{figure}[!h]
\begin{center}
\hspace{10.0truecm}
\includegraphics[width=0.46\textwidth]{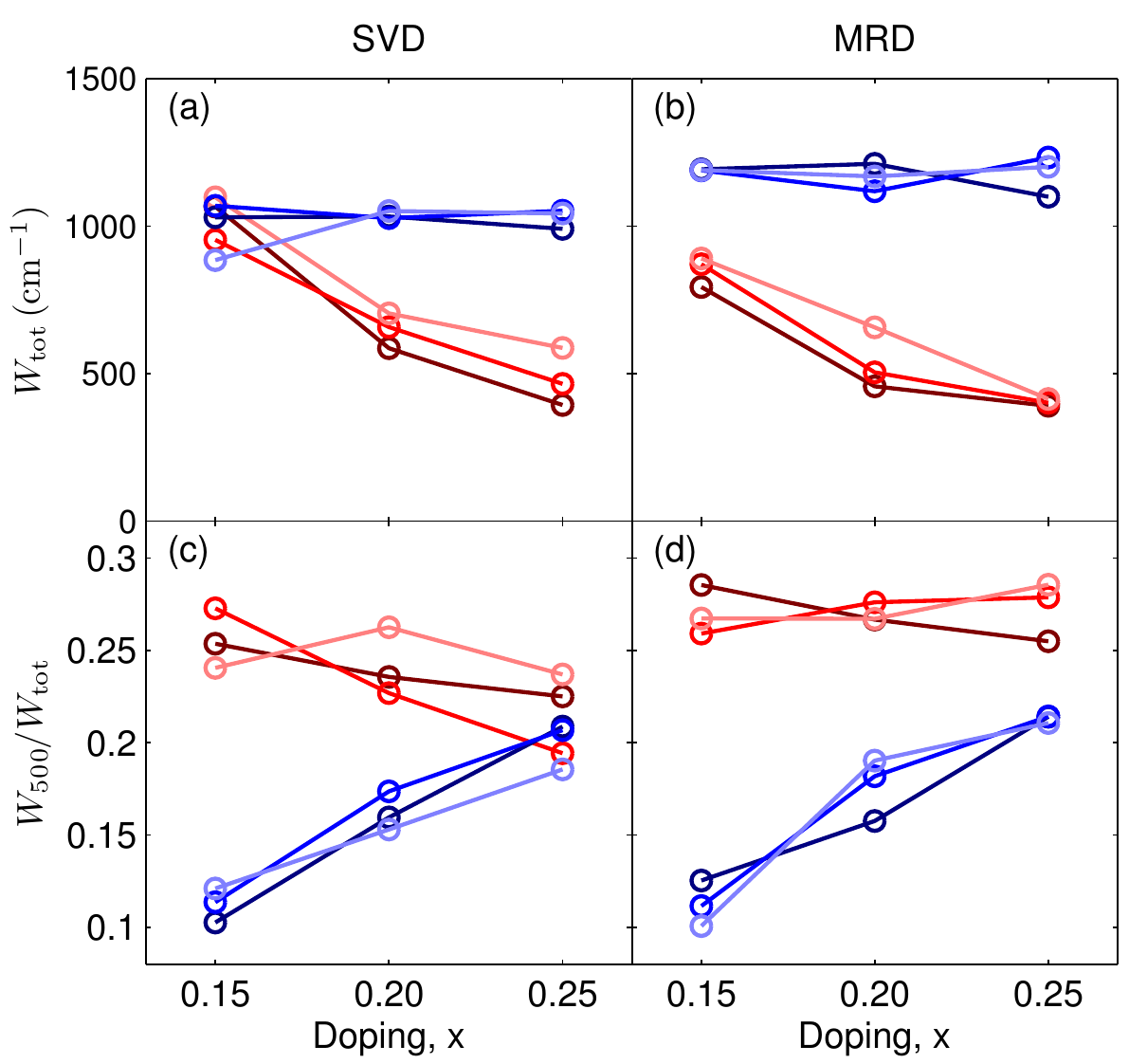}
\caption{(Color online) Panels (a) and (b): Total area (weight) $W_{\text{tot}}=\int \dd 
z\;\alpha^2F(z)$ under the glue functions as a function of doping $x$. Panels 
(c) and (d): weight fraction below 500\,\cmi, computed as 
$W_{500}/W_{\text{tot}}$ with $W_{500}=\int_0^{500} \dd z\;\alpha^2F(z)$. SVD 
results are shown in panels (a) and (c), while MRD results are shown in 
panels (b) and (d). Shades of red: $B_{1g}$ channel. Shades of blue: $B_{2g}$ 
channel. Temperature increases from dark to light colors.}
\label{fig:figure2}
\end{center}
\end{figure}

We can quantify these features by computing the total spectral weight $W_{\text{tot}}$ of the 
glue function and the fraction of the weight below a cutoff frequency $\bar{\omega}=500$\,\cmi, as
\begin{gather}
W_{\text{tot}} = \int_0^{\infty} \alpha^2F(z)\;\dd z\,,\nonumber\\
\frac{W_{\bar{\omega}}}{W_{\text{tot}}} = \frac{\int_0^{\bar{\omega}} 
\alpha^2F(z)\;\dd z}{\int_0^{\infty} \alpha^2F(z)\;\dd z}\,.\nonumber
\end{gather}
In particular, the ratio between $W_{\bar{\omega}}$ and $W_{\text{tot}}$ gives us the magnitude of the 
low-frequency structure with respect to the higher-frequency one. Results obtained by MRD and SVD are shown 
in Fig.\,\ref{fig:figure2}. At all doping level and all temperatures we find good quantitative agreement 
between the integrated quantities computed via MRD and SVD.
Data show little dependence on temperature, while the effect of doping is much stronger. As described before, 
the total area of the $B_{1g}$ channel decreases markedly as the doping increases, with the total spectral weight 
at $x=0.25$ being half the weight at $x=0.15$. On the other hand, the fraction of weight below 500\,\cmi\ 
appears to be roughly constant, suggesting that the two structures are proportionally suppressed.
The $B_{2g}$ channel glue function shows quite the opposite behavior. In fact, the total weight is only slightly 
dependent on temperature, while the fraction of weight below $\bar{\omega}=500$\,\cmi\ increases from $\sim10\%$ 
at $x=0.15$ to $\sim20\%$.

\section{Low-frequency analysis: Temperature and Doping dependence}
The previous analysis show little temperature-dependence of the global 
properties of the glue function. However, the low-frequency part of the Raman 
response function has a clear dependence on temperature. A low-frequency 
expansion of the glue function allows us to study the behavior of the glue 
function at frequencies $\omega \lesssim T$. As 
detailed in Appendix\,\ref{sect:appendix_lowTemp_1}, the relation between the 
Raman response function $\chi''$ and the memory function $M$ can be approximated 
for finite temperature and small frequencies as
\begin{equation}\label{eq:lowFreq_1}
\chi''(\omega) = \chi_0\left[ \frac{\omega}{M''_0}+\mathcal{O}(\omega^3) \right]\,.
\end{equation}

Therefore, the low-frequency behavior of the response function is governed by 
the imaginary part of the memory function at zero frequency, $M''_0$. 
Interestingly, $M''_0$ can be directly connected to the low-frequency behavior 
of the glue function $\alpha^2F(z)$. As detailed in 
Appendix\,\ref{sect:appendix_lowTemp_2}, $M''_0$ is proportional to an 
``effective'' slope $s_{\text{eff}}$ of the glue function, obtained  by a 
weighted average of $\alpha^2F(z)/z$ with a fast decaying weight, similar to 
$\exp(-z/T)$. 
The relationship between $s_{\text{eff}}$ and $M''_0$ is given by
\begin{equation}\label{eq:lowFreq_3}
s_{\text{eff}} = \frac{3 M''_0}{4 \pi^3 T^2}\,.
\end{equation}
If the glue function is approximately linear for $z \lesssim T$, 
then $s_{\text{eff}}$ is approximately equal to its slope.
On the other hand, if the glue function has two peaks, one for $z \lesssim T$ and the other one in $z \gtrsim T$,
the quantity $s_{\text{eff}}$ is related to the magnitude and steepness of the first peak only, similarly to the ratio 
$W_{\bar{\omega}} / W_{\text{tot}}$ discussed before. 
As already discussed, while the derivative of the glue function 
cannot be reliably estimated from the fitted glued function without some kind of 
smoothening procedure, integrated quantities, like $s_{\text{eff}}$ in \eqref{eq:lowFreq_3},  are much better suited for the 
analysis of the glue function properties.

We find that $M''_0$ displays a clear temperature trend, which is well described as a $T^{1/2}$ 
scaling for both the channels. Furthermore, the two channels have a marked 
opposite dependence on the doping. If we use a simple linear model $M''_0 
\propto (a+bx)$ to include the effect of the doping, we obtain the following 
expressions: %
\begin{align}
 M''_0(T,x,B_{1g}) &\approx (190-570\, x)\cdot \sqrt{T} ~\text{cm}^{-1}\text{K}^{-1/2}\nonumber ~,\\ \label{eq:lowFreq_4}
 M''_0(T,x,B_{2g}) &\approx (300\, x)\cdot \sqrt{T} ~\text{cm}^{-1}\text{K}^{-1/2}~.
\end{align}
These scalings reflect into similar scalings for the slope of the Raman response at zero frequency and effective slopes 
$s_{\text{eff}}$ of the glue functions (see Appendix~\ref{sect:appendix_lowTemp} and figures therein for more details).
These results are in agreement with the analysis shown in the previous section. The total weight $W_{tot}$ in the 
$B_{1g}$ channel decreases almost linearly with the doping, but not $W_{500}/W_{tot}$, therefore reducing $s_{\text{eff}}$. 
On the other hand, in the $B_{2g}$ channel, the total weight is constant, but $W_{500}/W_{tot}$ increases with $x$; as the 
low-frequency peak ``drains'' spectral weight from the second one, the slope $s_{\text{eff}}$ also increases.

\begin{figure}[b]
\includegraphics[width=0.44\textwidth]{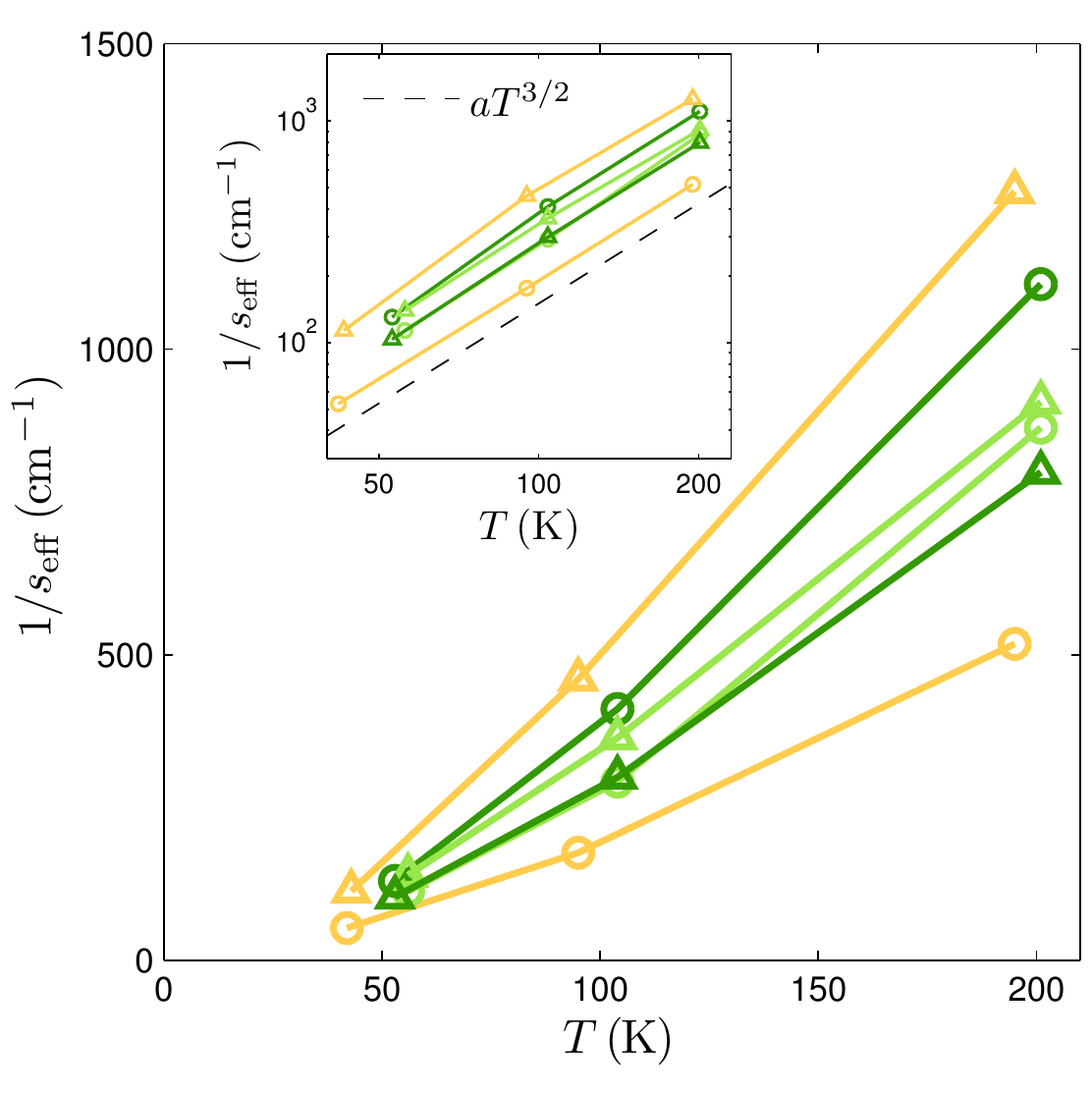} \caption{(Color online) Main panel: Temperature 
dependence of the 
inverse of the effective slope of the glue function at low frequency, $1/s_{\text{eff}}$, for different channels 
($B_{1g}$, circles; $B_{2g}$, triangles) and doping levels ($x=0.15$, yellow; $x=0.20$, pale green; $x=0.25$, dark 
green). Here, $s_{\text{eff}}$ has been computed from the $M_0''$ fitted using the SVD method. Similar results are
obtained with the MRD. Inset: The same data in a log--log scale, highlighting a seemingly $T^{3/2}$ behavior (dashed line).}
\label{fig9}
\end{figure}

\section{Discussion and Conclusions}
The Raman spectra in the two, B$_{1g}$ and B$_{2g}$, channels are markedly 
different: the $B_{1g}$ spectra have a rounded rise at low frequency until a new 
rounded growth takes place above 2000 cm$^{-1}$, while the $B_{2g}$ spectra have 
a flattish shape at low/intermediate frequencies until again a rounded growth 
takes place above 2000\,cm$^{-1}$. This difference naturally stems from 
different physical mechanisms ruling the low/intermediate-frequency scattering 
mechanisms. According to the most popular scenarios of superconducting cuprates, 
spin and charge collective excitations are the most likely candidates as 
mediators of scattering. In a previous work\cite{Caprara2011} we noticed that 
the different direction and magnitude of the characteristic wavevectors of these 
excitations produce different contributions in the two channels of the Raman 
spectra. Specifically, it was found that charge fluctuations with characteristic 
wavevectors $\qvec_c \approx(\pm \pi/2,0), (0,\pm \pi/2)$ (in reciprocal lattice 
units) are more visible in the low-frequency spectra in the $B_{2g}$ channel. On 
the contrary, spin fluctuations with characteristic wavevectors $\qvec_s 
\approx(\pm \pi \mp x,\pm \pi), (\pm \pi,\pm \pi \mp x)$ (in reciprocal lattice 
units) contribute more in the low-frequency spectra of the $B_{1g}$ channel. At 
higher frequencies both modes contribute to both channels. While the previous 
analysis assumed a phenomenological quantum critical form of the spin and charge 
excitations leading to analytic but obviously model-dependent glue functions, 
our main goal here is to put under scrutiny the previous results with an 
unbiased general analysis. Therefore within a memory-function approach we 
extracted the glue function directly form the Raman spectra without any 
assumption. To this purpose, we used two inversion methods: the SVD and the MRD. 
The overall shapes of the extracted glue functions are similar with both methods 
(see Fig.\,\ref{fig:figure1}): in both channels they show a rather narrow 
low-frequency peak below 500--1000\,cm$^{-1}$ and a broad hump up to 
3000--4000\,cm$^{-1}$. Despite the similarity of the two channels, a closer 
inspection shows that the doping dependency is just opposite, with the $B_{1g}$ 
glue function rapidly decreasing with doping above the optimal, while the glue 
function extracted from the $B_{2g}$ spectra stays roughly constant. More than 
this, one can notice that the frequency above which the $B_{2g}$ glue functions 
vanishes shrinks passing from the 3000--4000\,cm$^{-1}$ range, at $x=0.15$, to 
about 2000\,cm$^{-1}$, when $x=0.25$. This points to the natural interpretation 
that spin waves underly the $B_{1g}$ glue function at all frequencies, while 
charge excitations are responsible for the low/intermediate-frequency scattering 
in the $B_{2g}$ channel. In this channel spin waves also contribute, but mostly 
at high frequencies. When doping suppresses the spin-mediated scattering, the 
overall $B_{1g}$ glue decreases as well as the high-frequency part of the 
$B_{2g}$. The marked difference of the doping dependence in the two channels is 
also visible from both the low-frequency and total spectral weights of the glue 
functions (see Fig.\,\ref{fig:figure2}). A marked increase of the low-energy 
excitations involved in the $B_{2g}$ channel is also clear form Fig. 
\,\ref{fig:figure2}(c,d).

The glue functions extracted here are in overall agreement with those obtained 
from other techniques. A peaked feature below 1000 cm$^{-1}$ is also present in 
optical experiment \cite{vanheumen,dalconte} and in low-energy spin excitations 
revealed by inelastic neutron scattering (INS)\cite{wakimoto1,wakimoto2}. At 
higher energies (up to 0.3-0.4 eV) broad humps are found in optical spectra like 
those obtained within our analysis, which can be identified as due to spin 
excitations both from neutron scattering\cite{coldea} and from resonant 
inelastic X-ray scattering (RIXS) \cite{braicovich,dean}. At the same time there 
is increasing evidence from both NMR\cite{jullien} and 
RIXS\cite{ghiringhelli,chang,comin,letacon} that CDW are ubiquitous low-energy 
excitations in cuprates. It is therefore natural to identify them with the 
excitations involved in the $B_{2g}$ channel.

Our results also display a general agreement with  theoretical\cite{schachinger} 
and experimental\cite{he_prl13}, ARPES results in 
Bi$_2$Sr$_2$CaCu$_2$O$_{8+\delta}$ samples. In particular there is full 
consistency concerning the coexistence of two distinct collective modes and with 
the softening of the mode(s) near the ``hot'' antinodal region. It is noticeable 
that this agreement is present despite the obvious differences between the Raman 
response and the electron self-energy. In particular the former is resolved with 
respect to the electron momentum, while it is integrated over all the momenta of 
the bosons dressing the single-particle fermionic propagator. On the other hand, 
the Raman response obeys some (partial) cancellations when the boson momenta and 
the Raman form factors interplay\cite{Caprara2011}. From Ref. \onlinecite{he_prl13} one sees that 
the mode, which softens in the antinodal region, gets more strongly coupled in 
the underdoped region and could be identified with our spin-fluctuation mode. 
The (momentum averaged) energy of the other mode stays more constant over the 
electronic momentum space and it acquires  relatively more weight increasing the 
doping above optimal doping. This leads to the identification of this mode with 
our mixed phonon-CDW mode, which becomes prominent when doping increases above 
the optimal level. Concerning the temperature dependence, INS find below $T_c$  
a sharpening of the low-energy part of the spin-fluctuation spectra (see, {\it e.g.},\onlinecite{fong}), accounting 
for the clear observation of this mode in the ARPES spectra 
(Ref.s \onlinecite{schachinger, he_prl13}) below $T_c$ (the contribution of the other 
broad featureless, nearly temperature independent part of spin fluctuations 
being naturally buried in the broad incoherent parts of the electron 
self-energies). The softening of the charge modes extracted (mainly) from the 
Raman ($B_{2g}$) spectra is only expected to occur near the antinodal region and 
therefore in ARPES may be overshadowed by the softening of the other spin mode. 
Our selectivity in momentum allows instead to disentangle the contributions of 
the modes to the different Raman channels and to detect a substantial softening 
of the charge mode as well. 

Our work focuses on temperatures above $T_c$ and finds a distinct temperature behavior 
for the overall and the low-energy parts of the spectra. Specifically, we notice a rather weak 
dependence of the overall spectra, which becomes more pronounced in the 
low-frequency weight. This indicates that the excitations might have a 
low-frequency component with a marked temperature dependence. Indeed, as 
explained in Sec. IV, exploiting analytic low-frequency expansions of the memory 
function and of the glue functions extracted, e.g., via the SVD method, we are 
able to identify a substantial temperature dependence of the the imaginary part 
of the memory function at zero frequency. This is related [see Eq. 
(\ref{eq:lowFreq_1})] to the slope of the Raman response near $\omega=0$. In 
turn, this is related to the effective slope of the glue function 
$s_{\text{eff}}$ defined in Eq. (\ref{eq:lowFreq_3}), which is a rather robust 
quantity extracted from a weighted average of $\alpha^2F(z)/z$ that is only 
weakly sensitive on the oscillatory character of the SVD fitted glue functions. 
These strong temperature dependencies at low frequency are suggestive of some 
form of critical behavior of the mediating excitations. If for a while we borrow 
the analytic expressions of the glue functions calculated in 
Ref.\,\onlinecite{Caprara2011}, we find  $m/g \approx 1/s_{\text{eff}}$, where 
$m$ is the mass of a collective mode (i.e., the minimum energy needed to excite 
it) and is proportional to the square of the inverse correlation length of the 
corresponding fluctuations, while $g$ represents the coupling between fermionic 
quasiparticles and the collective modes. Thus, although in this work we 
purposely avoided any assumption on the mediator, we interestingly find that our 
results support a marked temperature dependence of its mass, as it is seen in 
Fig.\,\ref{fig9}.

Within a quantum critical scenario, the masses are expect to
vary linearly with $T$ in the quantum critical regime, and saturate in the quantum disordered regime.
Of course drawing any conclusion on this dependence with three temperatures only is out of question, but a 
clear indication of nearly critical mediators, with a certain tendency of the mass to flatten as a function of 
temperature at low $T$, can still be obtained.

\acknowledgments 
L.F. acknowledges the Spanish Ministerio de Econom\'ia y Competitividad (MINECO) 
through Grant No. FIS2011-29689. L.F. and M.M. acknowledge the Sapienza Universit\`a di Roma for 
financial support. S.C. and  M.G. acknowledge financial support form the Sapienza Universit\`a di Roma 
with the Project Awards n. C26H13KZS9.

\appendix
\setcounter{figure}{0}
\makeatletter
\renewcommand{\thefigure}{A\@arabic\c@figure}
\makeatother

\section{MULTI-RECTANGLE DECOMPOSITION} \label{sect:binning_appendix}
\label{sez:CT}

In this case the strategy to minimize the functional in Eq.\,(\ref{eq:Delta2}) is 
based on a histogram decomposition of the glue function, the fitting parameters 
being the heights of the histograms, in the $N$ different frequency intervals 
(bins) in which the frequency axis is partitioned, and the prefactor $\chi_0$ 
(that also include all multiplicative parameters needed to match the 
theoretically calculated response with the measured one). The fitting procedure 
searches for a local minimum, beginning from a starting guess, and using the 
direct search algorithm of Hooke and Jeeves.\cite{rif1,rif2,rif3,rif4} 

Using as starting point Eq.~\eqref{eq:integralEquation}, we divide the integration 
range into $N$ non-overlapping intervals as follows:
\begin{equation}
\label{app1}
M''(\omega)= \sum_{\alpha=1}^N \int_{a_\alpha}^{b_\alpha}\dd z \, K(\omega,z)\alpha^2F(z)\,,
\end{equation}
where $j=1,\ldots,N$, and to avoid confusion with the frequencies $\omega_i$ at 
which the Raman response is experimentally measured, we called here $a_\alpha$ and 
$b_\alpha$ the frequencies identifying the $\alpha$-th bin, with $a_\alpha<a_{\alpha+1}=b_\alpha$. In 
every integration interval of Eq.\,(\ref{app1}), the glue function  is to be considered a 
constant (the height of the bin) and can be taken out the integral and 
considered as a multiplicative factor indicated with $c_\alpha$, so we can 
rewrite the equation (\ref{app1}) in a matrix form 
\begin{equation}
\label{3_45_2}
M''(\omega_i)=\sum_{\alpha=1}^N c_\alpha \Delta M''_{i,\alpha},
\end{equation}
with
\[
\Delta M''_{i,\alpha} \equiv \int_{a_\alpha}^{b_\alpha}\dd z \,K(\omega_i,z)\,.
\]
The integration domain in Eq. (\ref{eq:integralEquation}) extends between $0$ 
and $\infty$. In Eq.\,(\ref{app1}) we have taken 
$b_N=8000$\,cm$^{-1}=z_{\mathrm{max}}$ assuming that the integrand vanishes at 
higher frequencies. The minimum frequency (acting as a natural cutoff within our
procedure) was typically taken as $a_1=10$\,cm$^{-1}=z_{\mathrm{min}}$.
Rather than adopting a homogeneous mesh, with constant 
$b_\alpha-a_\alpha$, we adopted a logarithmic mesh, with constant 
$b_\alpha/a_\alpha$, to enhance the sensitivity
at low frequency. The value of the integral in the bin with extremes $[a,b]$ 
appearing in Eq.\,(\ref{app1}), with $a,b>0$, is 
\begin{eqnarray*}
\Delta M''(\omega) &=& 2T\bar\omega^{-1}\alpha^2F(a)\Big\{ 2\bar\omega\ln \frac{\sinh(\bar b)}{\sinh(\bar a)}\nonumber\\
&+&\Xi(\bar a+\bar\omega)-\Xi(\bar a-\bar\omega)-\Xi(\bar b+\bar\omega)
+\Xi(\bar b-\bar\omega)\nonumber \\
&-&\sum_{k=1}^\infty \frac{1}{2k^2}\left[\Lambda(2k\bar a+2k\bar\omega)-\Lambda(2k\bar a-2k\bar\omega)\right]\nonumber \\
&+&\sum_{k=1}^\infty \frac{1}{2k^2}\left[\Lambda(2k\bar b+2k\bar\omega)-\Lambda(2k\bar b-2k\bar\omega)\right]
\Big\},\nonumber
\end{eqnarray*}
where $\bar a\equiv a/(2T)$, $\bar b\equiv b/(2T)$, 
$\bar\omega\equiv\omega/(2T)$, $\Xi(x)\equiv x[\ln(2\sinh|x|)-\frac{1}{2}|x|]$, 
$\Lambda(x)\equiv\mathrm{e}^{-|x|}\mathrm{sign}(x)$, and the last two lines 
contain rapidly convergent series. For numerical reasons, we approximated 
$\ln(\sinh|x|)\approx|x|-\ln(2)$ when $|x|>40$.

Once the imaginary part of the memory function is obtained, the real part is 
achieved through the KK transformation 
\begin{equation}
\label{app23}
M'(\omega) = \frac{1}{\pi}P\int_{-\infty}^{+\infty} \dd z \frac{M''(z)}{z-\omega}.
\end{equation}
Since the $M''(\omega)$ function is known at discrete points, to calculate the 
integral (\ref{app23}) the integrand function is replaced with a continuous 
broken line obtained by joining the heights of the bin with straight lines. So 
the analytical form of $M''(\omega)$ to be included in (\ref{app23}) will look 
like $M''(z)=Az+B$, in which the coefficients $A$ and $B$ are calculated within 
each bin. To calculate the integral (\ref{app23}) we have subdivided the 
integration interval in subintervals which are delimited by the same point where 
$M''(\omega)$ is known. Apparently, we can have two cases: $\omega$ is or is not 
an extreme of integration. In the first case the results of the integral 
(\ref{app23}) for the subinterval delimited by, say, $[\omega_1,\omega_2]$ is 
straightforwardly 
\[
A(\omega_2-\omega_1)+(A\omega+B)\ln \left|\frac{\omega_2-\omega}{\omega_1-\omega}\right|,
\]
while in the latter case two neighboring bins $[\omega_1,\omega-\xi]$ and 
$[\omega+\xi,\omega_2]$, treating the divergence in the sense of the principal 
value, have to be considered. When the results for the two neighboring bins are 
summed, the divergent part cancels and the same expression is obtained.

We have to remark that, in general, multiple choices of $\alpha^2F$ yield
the same memory function $M''$, due to the presence of nonzero solutions to the 
matrix equation $\sum_{\alpha=1}^{N} \Delta M''_{i,\alpha} c_\alpha = 0$.
It is therefore impossible to fit an unique glue function $\alpha^2F$
without imposing additional constraints (see, e.g., Ref.\,\onlinecite{rif5}). 
We restrict the possible glue functions by imposing the following constraints:
\begin{enumerate}
\item $\alpha^2F(\omega) \ge 0$, and therefore $c_\alpha \ge 0$;
\item The initial guess for the glue function in the minimization procedure
has vanishing spectral weight for $\omega>4000$\,cm$^{-1}$, 
assuming that the collective modes live at lower frequencies.
\end{enumerate}
Such constraints do not completely avoid the uncertainty about the glue 
function. However, the relevant features of the calculated glue function are 
robust enough, upon varying the doping and the temperature, as discussed in 
Sec.\,\ref{sect:fit_results}.

We performed our calculations comparing the results obtained with $N=$8,12, and 25
bins, to ensure that the main features of the glue function were robust with
respect to variations in the MRD scheme.

\setcounter{figure}{0}
\makeatletter
\renewcommand{\thefigure}{B\@arabic\c@figure}
\makeatother

\section{SINGULAR VECTOR DECOMPOSITION} \label{sect:SVD_appendix}
In this section we describe in a detailed way the SVD-based fitting procedure. 
The first step is to discretize the kernel $K(\omega,z)$ in 
Eq.\,\eqref{eq:kernelDefinition}. This is done by using two (possibly different) 
meshes $\{w_i\}$ and $\{z_j\}$. Logarithmic meshes, or combinations of linear 
(at small frequencies) and logarithmic (at large frequencies) meshes, can be 
used. The integration measure is associated to the kernel, so that 
Eq.\,\eqref{eq:integralEquation} becomes: 
\begin{equation}
M''_i =\sum_j K_{ij} \alpha^2F_j\,. \label{eq:discretized_integral_equation}
\end{equation}
The SVD, although computationally demanding, provides the best low-rank 
approximation of the kernel $K$, a result known as the Eckart--Young theorem. In 
this sense, it is the ideal tool to approximate solutions of integral equations. 
However, a na\"{i}ve application of the SVD to the kernel in
Eq.~\eqref{eq:discretized_integral_equation} produces basis vectors $\phi_{\alpha}$ with undesirable properties.
(1) On physical grounds, we require that the glue function goes to zero at large frequency (at least as $1/z$,
see Ref.\,\onlinecite{goetze1972_memfun}).
However, the integration kernel $K$ is long-ranged, in the sense that it is nonzero for large $\omega$ and $z$.
As a result, the singular vectors $\phi_\alpha(z)$ do not go to zero for large $z$,
making hard to constrain $\alpha^2F$ to be small at large frequencies.
(2) Even worse, the kernel $K(\omega,z)$ diverges as $\sim 1/z$ for $z\to 0$; therefore,
the SVD of the kernel produces basis vectors $\phi_{\alpha}$ which also divergent in 0.
This is in sharp constrast with the fact that the glue function should
go to zero at small frequencies in order to ensure the
convergence of the integral, Eq.~\eqref{eq:integralEquation}. 
As described in Appendix~\ref{sect:binning_appendix}, within the MRD approach
these two problems are easily solved by choosing bins with lower edges larger
than zero (thus providing a cutoff to the integral) and upper edges smaller than a maximum frequency 
$\sim 8000$\,\cmi. 
In the SVD context, the solution to these problems requires to expand a properly
regularized kernel instead of the original one; we will detail our procedures in the next sections.

\subsection{Large frequencies}
In order to perform numerically the integral in 
Eq.\,\eqref{eq:integralEquation}, some cutoff frequency $\Omega$ has to be 
introduced. Due to the long-range nature of the kernel, the $\phi$ 
eigenfunctions are strongly dependent on the cutoff. Therefore, the choice of 
the cutoff should be motivated by physical reasons, and we have to check that 
the results are only weakly dependent on the particular cutoff choice.

Let us note that the integration kernel $K(\omega,z)$ goes to zero for $z\gtrsim 
\omega + \mathcal{O} \left( e^{-(z-\omega)/T} \right)$ [see 
Eq.\,\eqref{eq:kernelDefinition}]. This means that high frequency ($z > \Omega$) 
components of the glue function do not contribute to the memory functions at low 
frequency ($\omega < \Omega$) and, vice versa, that only the low-frequency part 
of the glue function can be fitted from low-frequency data.

We can explicitly set to zero the high-frequency part of the glue function 
introducing a cutoff function $q(z)$ as
\begin{gather} \label{eq:cutoffDefinition_1}
\alpha^2 F(z) = q(z) f(z),
\end{gather}
where $q(z)$ is 1 at low frequencies and goes to zero at high frequencies, and 
$f$ does not diverge in the $z\to\infty$ limit. Many choices for the cutoff 
function $q(z)$ are possible, for instance a power law decay 
\begin{align} 
\label{eq:cutoffDefinition_2a}
 q(z) &= \frac{1}{1+(z/\Omega)^\gamma}\,,
\end{align}
or an exponential decay, e.g., 
\begin{align}
q(z) &= \frac 1 2 \left[ 1 + \tanh\left(\frac{\Omega-z}{\Delta}\right)\right],\nonumber
\end{align}
for suitable parameters $\Omega$, $\gamma$ and $\Delta$. We find that the 
results depend only weakly on the cutoff, as the support of the fitted glue 
functions is concentrated before 2000--3000\,\cmi. Therefore, we will write $M'' 
( \omega) = \int\dd z \; K(\omega,z) q(z) f(z)$ and fit $f(z)$ to the 
experimental data using the cut off kernel $K(\omega,z) q(z)$. After a function 
$f(z)$ has been determined, the corresponding glue function is obtained from 
Eq.\,\eqref{eq:cutoffDefinition_1}.

\subsection{Small frequencies}
The divergence in $z=0$ of the kernel implies that the right eigenfunctions 
(which are used as a basis function for the glue function) are divergent in 
zero. Since $\alpha^2F(0)=0$, we want to approximate the glue function using a 
set of functions for which $\phi_\alpha(0)=0$. This can be accomplished by 
expanding a suitable regularized kernel instead of the original one. We can 
write:
\begin{equation} \label{eq:kernel_decomposition}
 K(\omega,z) = K_R(\omega,z) + K_D(z)~,
\end{equation}
where $K_R$ is regular in $z\to 0$, and $K_D(z)=4\pi T/z$. We can use the SVD 
decomposition for the regular part of the kernel, and study the effect of the 
second term on the solution. Of course, $K_D(z)$ is still divergent, but the 
right eigenfunctions of $K_R$ go to zero fast enough to regularize the integral.

\subsection{Rewriting the integral equation}
Putting together the cutoff definition, Eq.~\eqref{eq:cutoffDefinition_1}, and the decomposition into
regular and divergent part, Eq.~\eqref{eq:kernel_decomposition}, we are able to
recast the initial equation Eq.~\eqref{eq:integralEquation} in the following form:
\begin{align} 
\label{eq:integralEquation_afterTreatement}
 M''(\omega) = & \int \dd z\; [K_R(\omega,z)q(z)] f(z) \nonumber\\
 &+ \int \dd z\; [K_D(z)q(z)] f(z).
\end{align}
We then perform the SVD on the new kernel $K'(\omega,z) \equiv [K_R(\omega,z) q(z)]$, 
rather than on the original kernel $K(\omega,z)$, as follows:
\begin{equation}
\label{eq:appendix_SVD_expansion}
 K_R(\omega,z)q(z) \approx \sum_{\alpha=1}^N \sigma_\alpha \psi_\alpha(\omega) \phi_\alpha(z),
\end{equation}
where $N$ is the number of singular vectors we use to approximate the kernel.
We can hence rewrite Eq.\,\eqref{eq:integralEquation_afterTreatement} by expanding 
$f(z) = \sum_\alpha c_\alpha \phi_\alpha(z)$, obtaining
\[
 M''(z) = \sum_\alpha c_\alpha A_\alpha''(z) \,,
\]
where
\begin{align}
 A''_\alpha(\omega) &= \sigma_\alpha \psi_\alpha(\omega) + \delta \sigma_\alpha\,,\nonumber\\
 \delta \sigma_\alpha &= \int \dd z\; K_D(z)q(z)\phi_\alpha(z).\nonumber
\end{align} 
The real part of the memory function, $M'(z)$ can be obtained as a function of 
the coefficients $c_\alpha$ by taking the KK transform of the functions 
$A_\alpha''(\omega)$ (or, equivalently, of the $\psi_\alpha(\omega)$ functions). 
Eq.\,\eqref{eq:im_chi_and_memfun_expanded} is then obtained by plugging the 
expressions for $M'$ and $M''$ into Eq.\,\eqref{eq:im_chi_and_memfun}. 
\begin{figure} 
\centering 
\includegraphics[width=.485\textwidth]{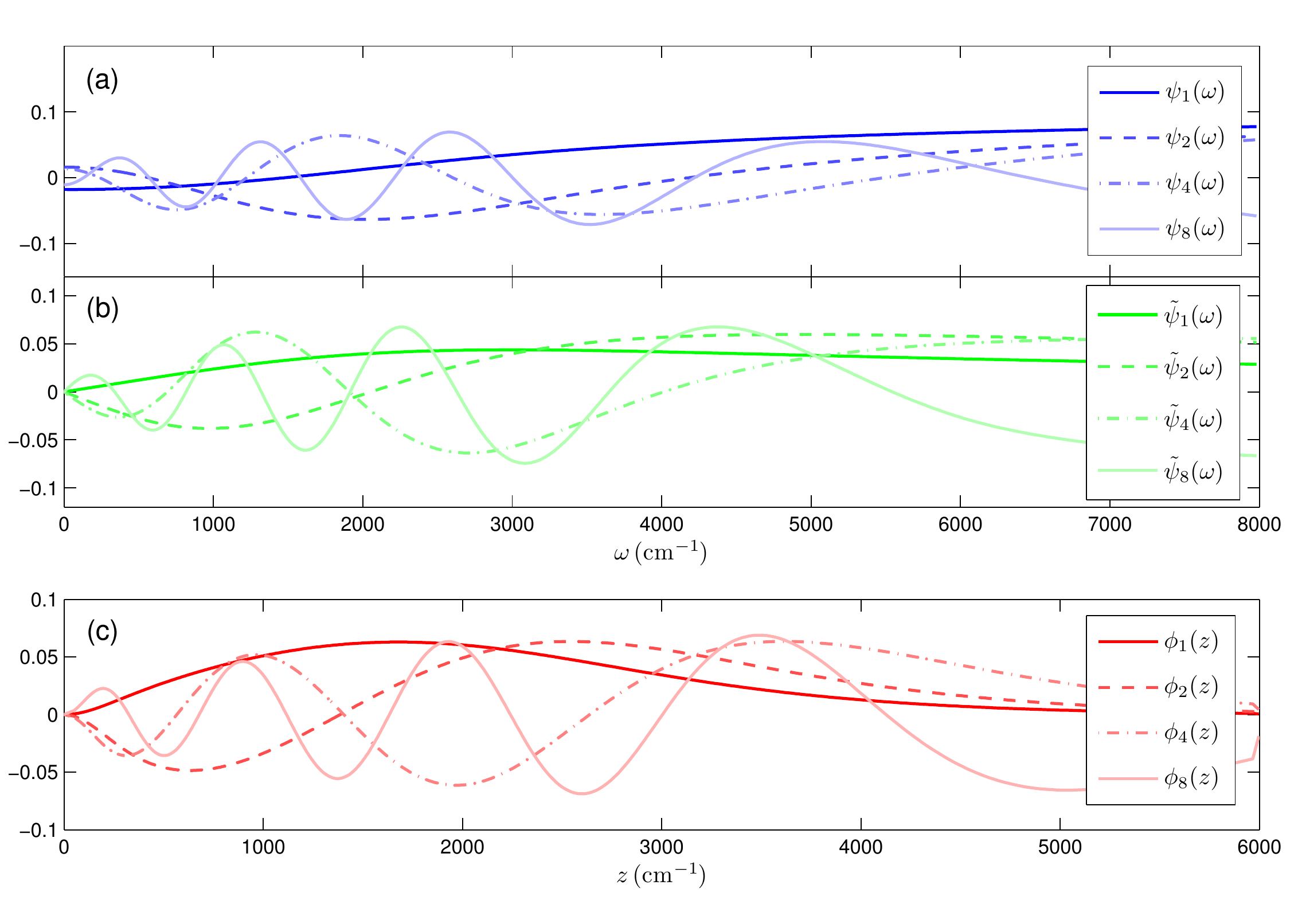}\vspace{-0.5cm} 
\caption{(Color online) Singular vectors from the SVD expansion of the 
regularized kernel, Eq.~\eqref{eq:appendix_SVD_expansion}, at $T=100$~K for some 
representative values of $\alpha=1,2,4,8$. (a) $\psi_\alpha(\omega)$ functions 
(b) Kramers-Kr\"onig transform $\tilde \psi_\alpha(\omega)$ of the 
$\psi_\alpha(\omega)$ functions (c) $\phi_\alpha(z)$ functions.
The eigenfunctions are qualitatively similar for all temperatures considered
in this work.
\label{fig:suppfig_SVD_functions}} \end{figure}

\subsection{Fitting procedure}
\label{sect:SVD_appendix_fitting}
A simultaneous fit of both $\chi_0$ and the expansion coefficients 
$\{c_\alpha\}$ is often unstable, with $\chi_0$ being pushed towards very large 
values. Therefore, we decided to fit the coefficients $c_\alpha$ at fixed 
$\chi_0$, and to systematically study the effect of different $\chi_0$. As 
explained in the main text, the coefficients $\{c_\alpha \}$ which fix the glue 
functions are obtained by the minimization of a function 
$\Delta^2(\{c_\alpha\})$, i.e., the square distance $\Delta^2$ between the 
theoretical response function and the experimental one.

It is useful to introduce the primitive of the glue function $W(z) = \int_0^z 
\dd z' \alpha^2F(z')$, and $W_{\max}=W(z_{\max})$, with $z_{\max}$ being the 
largest frequency in the $z$-mesh. Therefore, the quantity $1-W(\bar 
z)/W(z_{\max})$ equals the fraction of the area under $\alpha^2F$ at frequencies 
larger than $\bar z$.

\begin{figure}
\centering
\includegraphics[width=.47\textwidth]{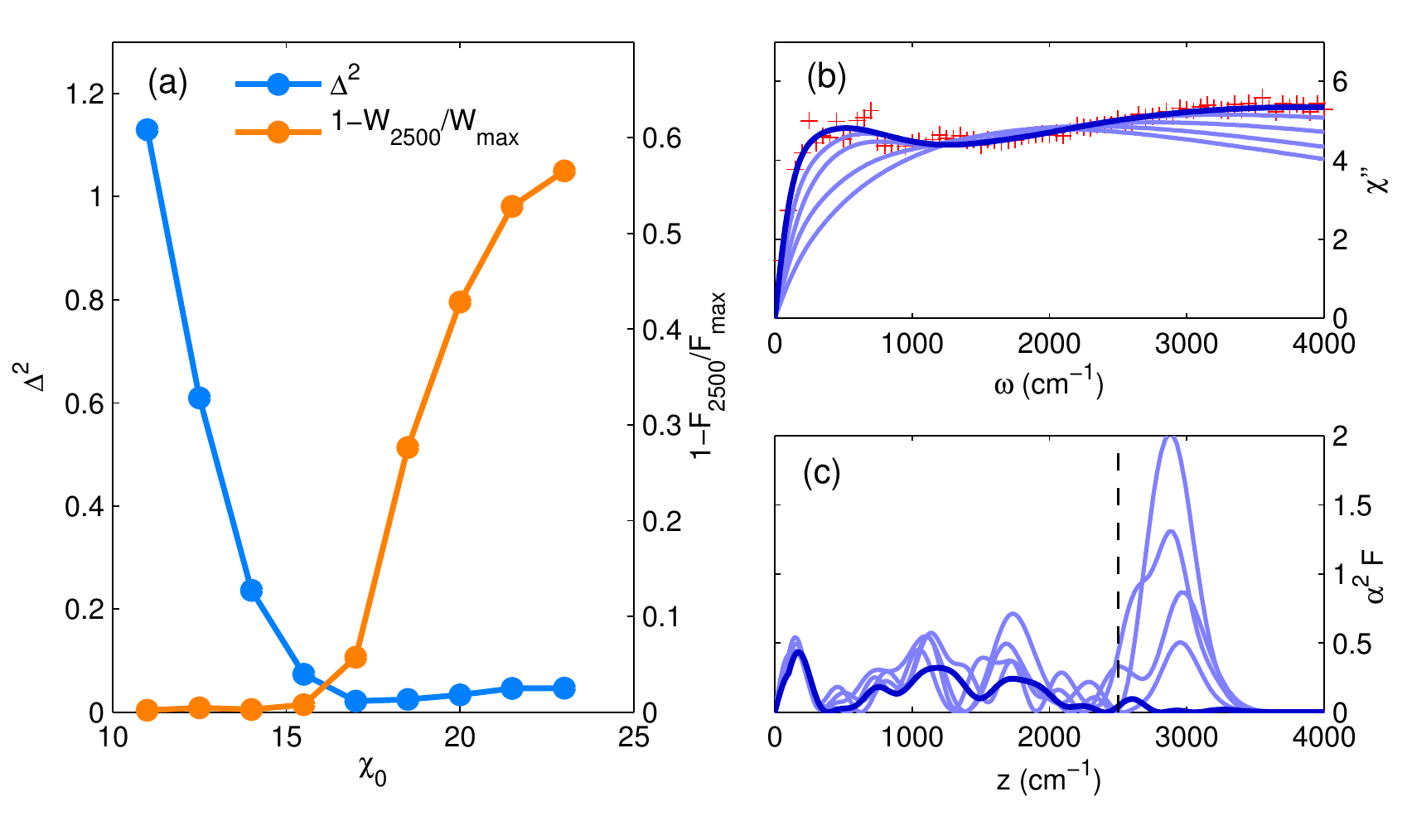}
\vspace{-0.5cm}
\caption{(Color online) (a) Fit of $\chi''$ using different $\chi_0$ values. We show the 
goodness of fit, as measured by $\Delta^2$ (see Eq.\,\eqref{eq:Delta2}), and the 
ratio between the area under the glue function for $\omega \ge 2500$\,\cmi\ to 
the total glue function weight $W_{\max}$. Best fit is obtained for $\chi_0 \sim 
17$. (b) Smaller values of $\chi_0$ do not yield a good fit, with $\chi_{th}''$ 
being systematically smaller than the experimental susceptibility. (c) The 
fitted glue function increases its total area, in particular at high 
frequencies, as $\chi_0$ is increased. The lines shown in panel (b) [or (c)] 
correspond to the best fit value $\chi_0=17$ (blue thick line) and the $\chi_0$ 
values shown in panel (a), smaller (or larger) than 17 (thin lines). All points 
have been computed using the same mesh ($N_w=536$, $N_z=507$) and cutoff 
(power-law, see Eq.\,\eqref{eq:cutoffDefinition_2a} with $\Omega=3000$\,\cmi\ 
and $\gamma=4$).\label{fig:fitGoodness_3}} 
\end{figure}

Fig.\,\ref{fig:fitGoodness_3} shows an example (data corresponds to the $B_{1g}$ 
channel, $T=105$\,K, doping $x=0.25$). As one can see in panels (a) and (b), 
small values of $\chi_0$ do not allow for a good fit, while for 
$\chi_0 \gtrsim 17$ the $\Delta^2$ is generally small. However, we see that using 
a large $\chi_0$ produces glue functions with a sizable spectral weight at high 
frequencies, as it is clear from panels (a) and (c). In this case the choice 
$\chi_0=17$ gives the best combination of fit quality and reduced spectral 
weight at high frequencies.

\begin{figure}
\centering
\vspace{0.2cm}
\includegraphics[width=.45\textwidth]{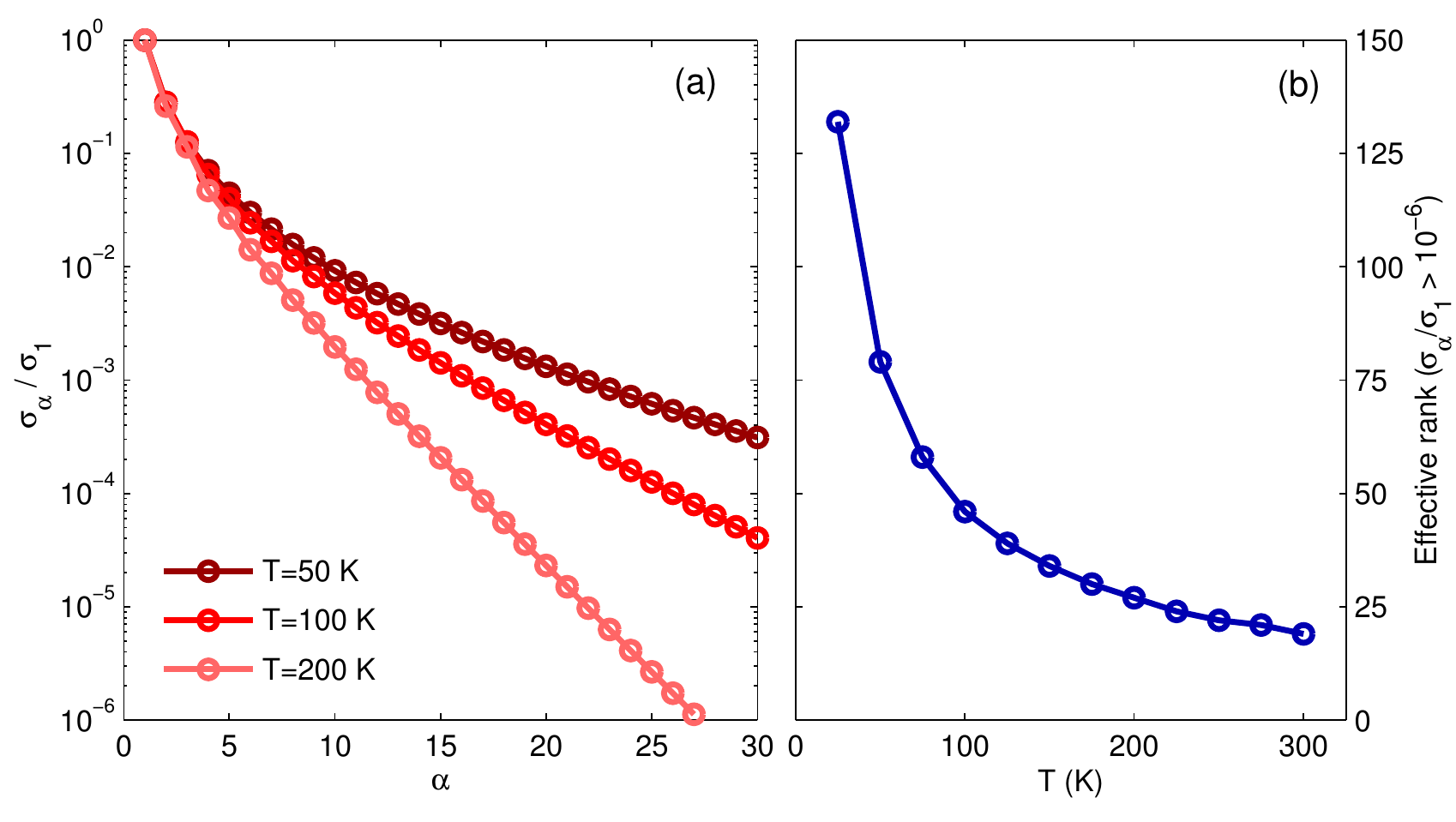}
\vspace{-0.5cm}
\caption{(Color online) (a) Magnitude of the singular values $\sigma_\alpha$ of the 
kernel $K_R(\omega,z)q(z)$, at three different temperatures, normalized by the 
value of the first SV. (b) Number of singular values $\sigma_\alpha$ larger than 
$10^{-6}\times \sigma_1$ as a function of temperature. In both panels the same 
mesh ($N_w=536$, $N_z=507$) and cutoff (power-law, see 
Eq.\,\eqref{eq:cutoffDefinition_2a} with $\Omega=3000$\,\cmi\ and $\gamma=4$) 
was used.
 \label{fig:rank}}
\end{figure}

We finally address the question of how to choose the number of singular vectors. 
The contribution of each singular vector $\phi_\alpha$ to the susceptibility is 
mediated by the singular values $\sigma_\alpha$. The number of singular vectors 
which have to be kept into account depends on how fast $\sigma_\alpha$ decreases 
with the index $\alpha$. As one can see in Fig.\,\ref{fig:rank}, the decay of 
the singular vectors strongly depends on temperature. At high temperature, the 
kernel can be approximated by just using a few terms of the SVD Hence, the 
number of singular vectors used to approximate the kernel should decrease with 
$T$, too. Moreover, we find that the convergence of the fit gets considerably 
worse if too many singular vectors are used. A reliable choice is using 40 SVs 
for $T\sim50$\,K, 25 SVs for $T\sim100$\,K, and 15 SVs for $T\sim200$\,K.

\setcounter{figure}{0}
\makeatletter
\renewcommand{\thefigure}{C\@arabic\c@figure}
\makeatother

\section{RELATING LOW-FREQUENCY BEHAVIOR OF RAMAN RESPONSES AND GLUE FUNCTIONS} 
\label{sect:appendix_lowTemp}
\subsection{Raman response and memory function} 
\label{sect:appendix_lowTemp_1}
\begin{figure}[!h]
\begin{center}
\hspace{10.0truecm}
\includegraphics[width=0.5\textwidth,clip=true,trim=.7cm 0cm 0cm 0cm]{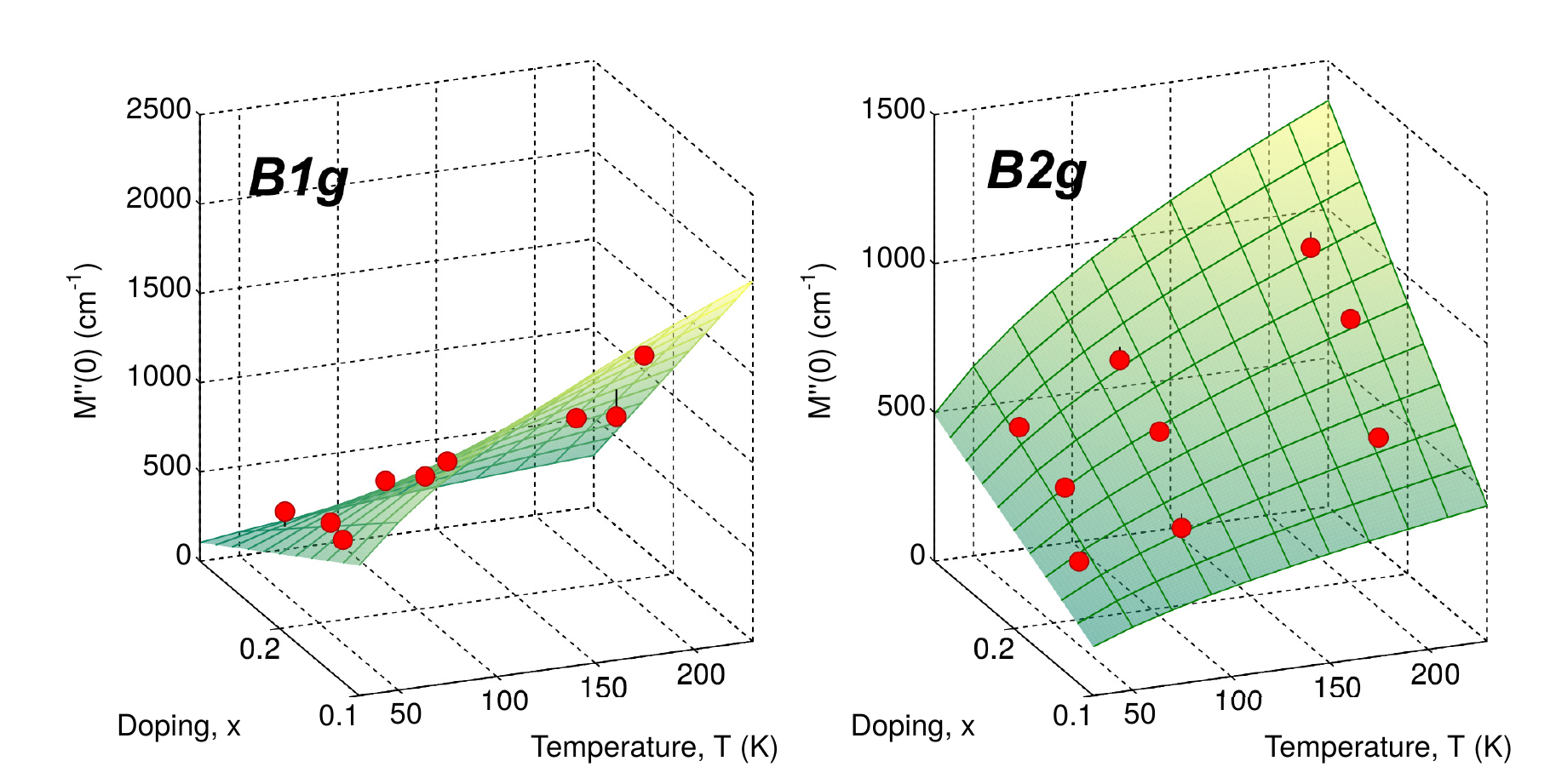}
\caption{(Color online) The red points are $M''_0$ using the SVD procedure, as a function of temperature and doping, for the 
two channels. The surfaces represent the fits $M''_0(T,x)=\sqrt T(a+bx)$. The surface parameters are $a=190$ and 
$b=-570$ for the $B_{1g}$ channel, while they are $a=0$ and $b=300$ for the $B_{2g}$ channel, where both $a$ and $b$ 
are in $\text{K}^{-1/2}\text{cm}^{-1}$ units.}
\label{fig:fig_M2}
\end{center}
\end{figure}

Some informations on the glue function can be directly extracted from the 
low-frequency behavior of the experimental Raman response function $\chi''$. In 
fact, let us expand both the real and imaginary parts of the memory function 
near zero as
\begin{align}
M'(\omega) &=M'_1 \omega+\mathcal{O}(\omega^3),\nonumber\\
M''(\omega) &=M''_0 +M''_2\omega^2 +\mathcal{O}(\omega^4),\nonumber
\end{align}
where we exploited the opposite parity of the functions $M'$ and $M''$.
Substituting into Eq.\,\eqref{eq:im_chi_and_memfun} we obtain 
\begin{equation} 
\label{eq:lowFreq_1_supp}
\frac{\chi''(\omega)}{\chi_0} = \frac{\omega}{M''_0}  
+\left( \frac{\omega}{M''_0} \right)^3 \left[(M'_1+1)^2 - M''_0 M''_2 \right] +\mathcal{O}(\omega^5).
\end{equation}
Therefore, the low-frequency behavior of $\chi''$ is controlled by the imaginary 
part of the memory function, computed at zero frequency. As described in the 
main text, $M''_0$ is found to have a marked dependence on temperature and 
doping, as described by Eq.\,\eqref{eq:lowFreq_4} and shown in Fig.\,\ref{fig:fig_M2}. 

\begin{figure}
\includegraphics[width=.45\textwidth]{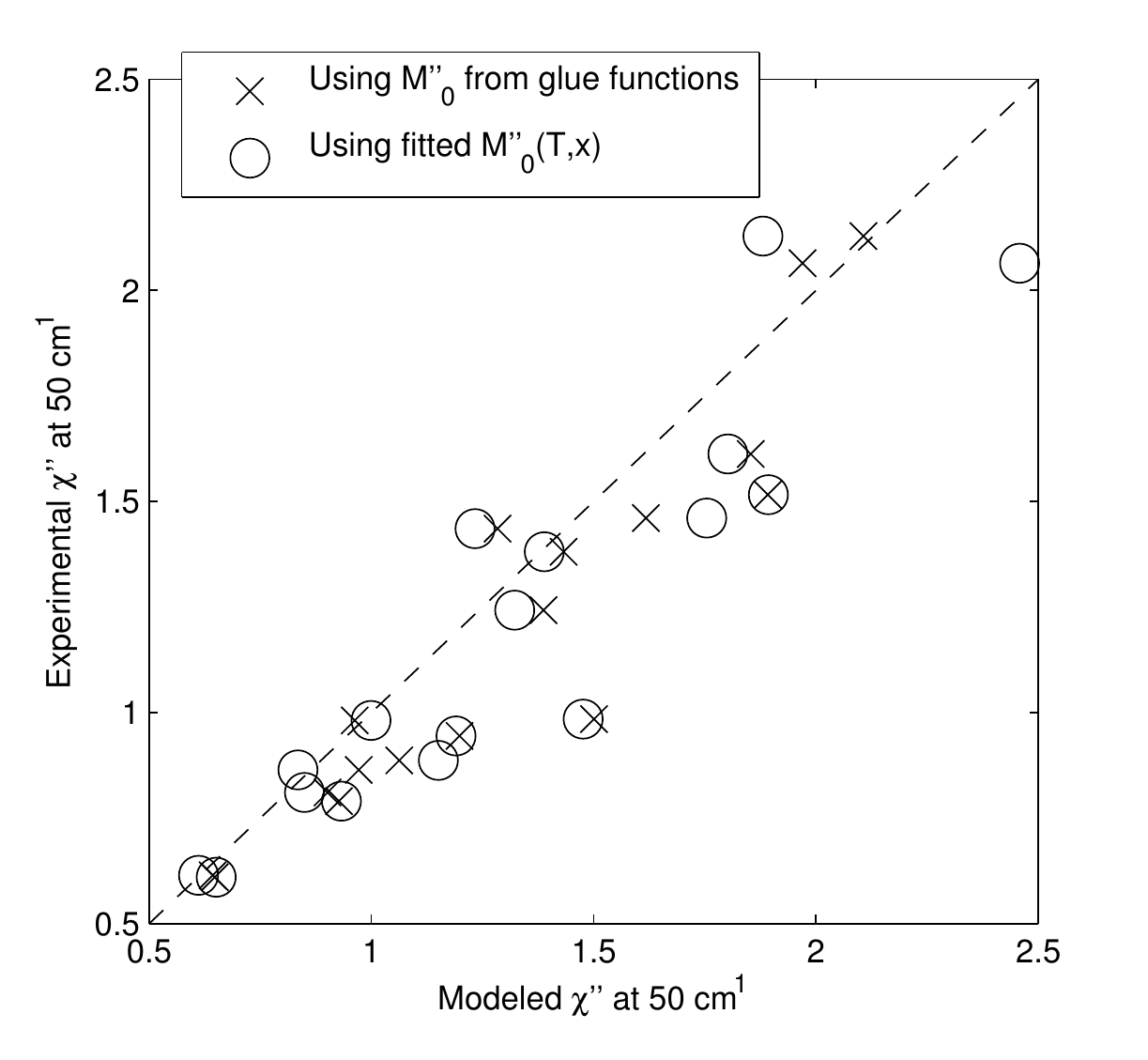}
\caption{(Color online) Comparison between the experimental value of the Raman response 
functions at $\omega=50$\,\cmi\ and the values obtained using the approximation 
$\chi''(\omega) \sim \chi_0\omega/M''_0$, with $M''_0$ being directly computed 
from the (SVD-) fitted glue functions of with the fitted models in 
Eq.\,\eqref{eq:lowFreq_4}. In both cases, $\chi_0$ is set to the values obtained 
during the fitting procedure. \label{fig:supp_fig_chi2_50}}
\end{figure}

The expansion Eq.\,\eqref{eq:lowFreq_1_supp} allows an estimate of the Raman response function 
$\chi''(\omega)$ in the low-frequency limit. Fig.\,\ref{fig:supp_fig_chi2_50} 
shows how the experimental $\chi'' (\omega)$ computed at $\omega=50$\,\cmi\ (the 
lowest nonzero frequency available for all the datasets) compares with the first 
order expansion $\chi_0 \omega/M''_0$, where $M''_0$ can be either the value 
computed by integrating the fitted glue functions, or the value extracted from 
the fits in Eq.\,\eqref{eq:lowFreq_4}. The agreement is far from perfect, 
although the correct trend is respected. We have to remark that the single 
experimental points used in the comparison can be strongly influenced both by 
the limited number of data points at low temperature, by random instrumental 
errors and/or by the presence of phonon excitations. As a consequence, the 
scaling results, Eq.\,\eqref{eq:lowFreq_4}, obtained using $M''_0$ (arising from 
global fits to the data) are much more consistent than the ones obtained with 
$\chi''(\omega)$ alone.

\begin{figure}
\includegraphics[width=0.46\textwidth]{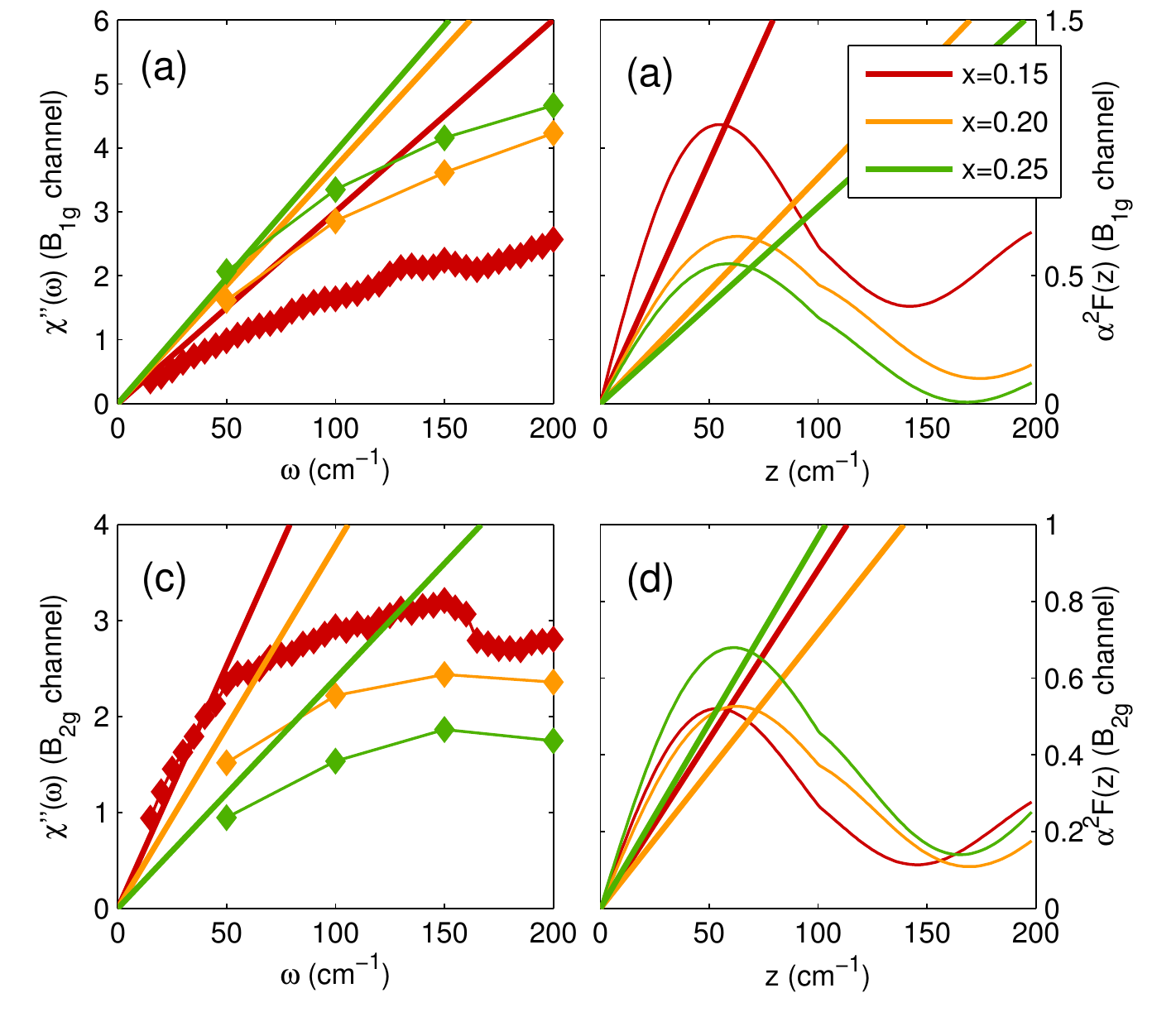} 
\caption{(Color online) Low frequency 
behavior of Raman response function $\chi''$ and glue functions $\alpha^2F$ at 
$T\sim 50$\,K. The $B_{1g}$ channel is shown in panels (a) and (b), while the 
$B_{2g}$ channel is shown in panels (c) and (d). The straight lines in panels 
(a) and (c) correspond to $\chi'' \sim \chi_0\omega/M''_0$, see 
Eq.\,\eqref{eq:lowFreq_1}; the lines in panels (b) and (d) correspond to 
$\alpha^2F(z) = s_{\text{eff}} z$, where $s_{\text{eff}}$ is the effective slope 
computed from $M''_0$, see Eq.\,\eqref{eq:lowFreq_3}. All the lines are computed 
with the $M''_0$ and the $\chi_0$ extracted using the SVD fitting procedure.}
\label{fig:fig_M2_slope}
\end{figure}

\subsection{Memory function and glue function} 
\label{sect:appendix_lowTemp_2}
Equation \eqref{eq:integralEquation} can be used to relate $M''_0$ and the glue 
function as follows 
\begin{align} 
M''_0 &= \int_0^\infty \dd z\; K(0,z) \alpha^2 F(z) \nonumber\\
&= 2\pi \int_0^\infty \dd z\; \frac{z}{2T}\text{cosech}^2 \left( \frac z {2T} \right) \alpha^2F(z)~.\nonumber
\end{align}
The kernel $K(0,z)$ has a $1/z$ divergence at $z=0$ and dies exponentially as 
$\exp(-z/T)$. Therefore, the support of $K(0,z)$ is essentially the interval $z 
\in [0,T]$, so that $M''_0$ is a measure of the spectral weight of the glue 
function at frequencies $\omega \lesssim T$. In particular, at low temperature, the slope 
at zero frequency of the glue function is the dominant contribution in the integral. This suggests 
the definition of the quantity
\begin{equation}
s_{\text{eff}} \equiv \frac{3M''_0}{4\pi^3T^2}\,. \label{eq:appendix_s_eff_definition}
\end{equation}
It is easy to show that, in the $T\to 0$ limit, $s_{\text{eff}}$ coincides with the slope 
$\dd \alpha^2F(z) / \dd z$ at $z=0$. This can be seen by expanding in series the glue function around $z=0$ 
as $\alpha^2F(z)=c_1 z + c_2 z^2 +c_3 z^3+\dots $. The coefficients $c_i$ depend in principle on both temperature, 
doping and channel. Then, we can compute the contribution of each term to $M''_0$ (this may not be a convergent 
series, it depends on how fast the coefficients $c_n$ decay with $n$; however, we assume that it is at least 
an asymptotic series for low enough $T$): 
\begin{align}
M''_0 & \sim \sum_{n\ge1} \int_0^\infty K(0,z) z^n\,\dd z \nonumber \\
&= \sum_{n\ge 1} 2\pi \left(2T\right)^{n+1} \int_0^\infty \text{cosech}^2(y)\, y^{n+1}\,\dd y  \nonumber  \,.
\end{align}
The function $J(x) = \int_0^\infty \text{cosech}^2(t)\, t^{x}\,\dd t$ is well defined for complex-valued arguments 
in the domain $Re(x)>1$, and can be expressed as 
$J(x) = 2^{1-x}\zeta(x)\Gamma(x+1)$, $Re(x)>1$,
where $\Gamma(x)$ and $\zeta(x)$ are Euler's gamma and Riemann's zeta functions, 
respectively. The first few values of the integral are $J(2)=\pi^2/6$, 
$J(3)=3\zeta(3)/2$ and $J(4) = \pi^4/30$. This leads to the result 
\begin{align}
M''_0 & \sim \sum_{n\ge 1} 2\pi \left(2T\right)^{n+1} J(n+1) \nonumber \\
 & \sim  4\pi T \sum_{n\ge1} c_{n} T^n \times \zeta(n+1)\Gamma(n+2) \nonumber\\
& \sim \frac{4\pi^3}{3} T^2 c_1 + 24\zeta(3)T^3 c_2 +\frac{16 \pi^5}{15} T^4 c_3+\dots \label{eq:M2_expansion}
\end{align}
Therefore, $s_{\text{eff}} \sim c_1$ if the first term dominates the series in Eq.~\eqref{eq:M2_expansion}. Using 
the known value of the integral $J(2)$, we see that another way to define $s_{\text{eff}}$, equivalent to 
Eq.~\eqref{eq:appendix_s_eff_definition}, is
\[
s_{\text{eff}} = 
\frac{\int_0^\infty \,\dd z \left[ \frac{\alpha^2F(z)}{z}\right]\left(\frac{z}{2T}\right)^2
\text{cosech}^2\left(\frac{z}{2T}\right)}
{\int_0^\infty \,\dd z \left(\frac{z}{2T}\right)^2\text{cosech}^2\left(\frac{z}{2T}\right)}.
\]
This expression represents an average of the quantity $\alpha^2F(z)/z$ with a 
fast-decaying weight $\approx (z/T)^2\exp(-z/T)$ for $z\gg T$.
We show in Fig.\,\ref{fig:fig_M2_slope} how the values 
$M''_0$ computed from the SVD-fitted glue functions can be used to estimate the 
low frequency behavior of the Raman response function $\chi''(\omega)$ [panels 
(a) and (c)], and the effective slope $s_{\text{eff}}$. We see that the 
extrapolated slope of $\chi''$ fits quite well with the experimental data, even 
when the data points are scarce. The effective slope of the glue function, 
instead, is systematically smaller than the slope of the fitted glue functions 
at $z=0$. This is due to the glue functions being concave. However, 
$s_{\text{eff}}$ appears to be roughly proportional to $\dd \alpha^2F / \dd z|_{z=0}$. 
In particular, this suggests that the dependence of $s_{\text{eff}}$ and 
$\dd \alpha^2F / \dd z |_{z=0}$ on doping are very similar.
Fig.\,\ref{fig9} in the main text reports the temperature dependence of the 
inverse effective slope $s_{\text {eff}}$, highlighting the strong temperature 
dependence of the low-frequency glue function, possibly related to the quantum 
critical behavior of the interaction mediator.\cite{Caprara2011} Although the 
plot is limited to three temperatures, this quantity seems to be quite 
compatible with a power-law dependence $s_{\text {eff}}^{-1} \sim T^{3/2}$, as 
it is seen from the inset of Fig. \ref{fig9}, calling for further investigation.

\bibliographystyle{prsty_no_etal}

\end{document}